\documentclass[12pt,prc]{revtex4}

\newcommand{\lb}{\left\langle}
\newcommand{\rb}{\right\rangle}
\newcommand{\be}{\begin{eqnarray}}
\newcommand{\ee}{\end{eqnarray}}
\newcommand{\tl}{\tilde}
\begin{document}

\title{Classical Computation of Elliptic Flow at Large Transverse Momentum} 

\author{Derek Teaney}
\affiliation{%
 Physics Department, Brookhaven National Laboratory,
                Upton, N.Y. 11973, U.S.A.
}
\author{Raju Venugopalan}
\affiliation{%
Physics Department and RIKEN-BNL Research Center, 
Brookhaven National Laboratory, Upton, N.Y. 11973,
U.S.A.
}

\date{\today}

\begin{abstract}
We compute the contribution of classical fields 
to the second Fourier coefficient ($v_2$) of the azimuthal gluon distribution 
at large transverse momentum in heavy ion collisions. 
We find that the classical contribution to 
the flow alone cannot account for the experimentally observed behavior 
of $v_2(p_t)$ at large transverse momentum $p_t$.

\end{abstract}


\maketitle

\section{Introduction}

The azimuthal anisotropy of the particles produced in heavy ion collisions is 
quantified by the Elliptic flow parameter $v2$. It is defined to be the 
second Fourier moment of the azimuthal distribution,  
\be
v_2 = \langle \cos(2\phi) \rangle
    \equiv { \int^{\pi}_{-\pi} d\phi \cos(2\phi) \int p_t dp_t
                {d^3N \over dy\, p_t\,dp_t\,d\phi}
        \over
       \int^{\pi}_{-\pi} d\phi  \int p_t dp_t
                {d^3N \over dy\,p_t\,dp_t\,d\phi}  }\,.
\label{eq1}
\ee
The RHIC (Relativistic Heavy Ion Collider) experiments have measured
the centrality dependence of $v_2$ and find a value for
$v_2$~\cite{Expt1} that is significantly larger for non-central
collisions than previous measurements at lower
energies~\cite{CERN}. Hydrodynamic models do a reasonable job of describing
the centrality and mass dependence of $v_2$~\cite{Hydro1,Hydro2}. The 
RHIC experiments have also measured $v_2(p_t)$, which is defined to be 
\be
v_2(p_t) = \frac{\int_{-\pi}^\pi d\phi\cos(2\phi) 
                      {d^3N \over dy\, p_t\,dp_t\,d\phi}}{\int_{-\pi}^\pi 
d\phi {d^3N \over dy\, p_t\,dp_t\,d\phi}} \, .
\ee
The data~\cite{Expt1,Expt2} show that, for a wide range of centralities,
$v_2(p_t)$ rises up to $~1.5$ GeV, flattens and then remains flat
up to the highest transverse momentum measured $p_t\approx 6$ GeV. 
For peripheral collisions (specifically 34\%--85\% central collisions), 
$v_2(p_t)$ can be as large as $~25\%$ at $p_t\sim 3$ GeV.

Hydrodynamic models 
describe the initial rise in $v_2(p_t)$ but overpredict the data at large 
$p_t$. The models predict a continued rise except perhaps in a Blast 
wave parametrization with extreme assumptions~\cite{Blast}. 
Partonic descriptions which 
account for the quenching of high $p_t$ ``jets'' as they traverse the hot QCD 
matter do not describe the data either~\cite{GVW}-unless  
unusually large partonic cross-sections are assumed~\cite{Molnar}.

In this note, we discuss the computation of $v_2$ in the 
Color Glass Condensate (CGC) approach to nuclear 
collisions~\cite{MV,KMW,KV}. At very high energies, the density of partons 
per unit area in the colliding nuclei becomes very large and saturates 
at a scale $Q_s\gg\Lambda_{QCD}$~\cite{GLR}. For RHIC, estimates give 
$Q_s\sim 1-2$ GeV~\cite{GyulassyMcLerran,Mueller,KV,KN}. 
The typical occupation number of partons is $1/\alpha_s(Q_s)>1$. Thus 
classical methods can be used to describe nuclear collisions at high energies.
The CGC approach has 
been used to describe successfully various aspects of the bulk 
properties of charged hadrons in heavy ion collisions, including the 
multiplicity~\cite{KV}, the centrality dependence of the 
multiplicity~\cite{KN}, the energy and rapidity dependence~\cite{KL} and 
the $p_t$ dependence of inclusive hadron spectra~\cite{Juergen}.

Elliptic flow, as defined in Eq.~1, is dominated by contributions from small 
momenta. In the classical approach, this involves computing diagrams to all 
orders in $Q_s/p_t$. This computation is very difficult to perform 
analytically. It can be computed numerically~\cite{Yasushi}. 
However, when the transverse momenta is large, $p_t\gg Q_s$, the 
classical equations can be linearized and the 
contribution to $v_2(p_t)$ can then be computed analytically. 
This computation is 
performed in this paper. We find that the contribution to $v_2(p_t)$ from 
classical gluon fields at large momenta is small and differs 
significantly from the 
measured distribution at high $p_t$. The origin of the empirically observed 
behavior of $v_2(p_t)$, at large $p_t$, must arise from a source outside 
the classical perturbative expansion. Interestingly, even though the 
contribution from high momenta is small, this contribution is generated 
at very early times.

The paper is organized as follows. In the following section, we set up the 
problem of computing the classical gluon distribution at large transverse 
momentum. In section 3, we perform the computation. A brief final section 
summarizes our results. Some details of the computation are relegated to 
an appendix.

\section{The Classical Gluon Distribution}

Within the framework of the McLerran-Venugopalan model,
Kovner, McLerran, and Weigert (KMW)~\cite{KMW} (see also Refs.~\cite{KR} and 
~\cite{GyulassyMcLerran}) computed the classical 
spectrum of hard gluons (with $p_t>Q_s$) radiated from a collision of two 
infinitely large nuclei with  uniform color charge distributions. In order to 
calculate the azimuthal distribution of the radiated gluons, we 
generalize this work to a finite nucleus with a non-uniform 
color charge distribution.  Gradients in the
charge distribution then give rise to  space-momentum correlations
of the radiated gluons. The typical momentum 
scale of the produced gluons is $\sim Q_{s} (x_t)$. A typical
spatial gradient is $\sim 1/R$. Therefore, the 
typical space-momentum correlation is of order $\sim (Q_{s}R)^{-1}$.
The final expression for $v_{2}$ is actually of order $\sim (Q_{s}R)^{-2}$ 
since elliptic flow responds to the quadrupole moment of the 
color charge distribution.

We first review the perturbative solutions 
to the classical Yang Mills equations in order to establish the notation.  
In  the McLerran-Venugopalan model, 
two nuclei (A and B) collide at very high energies.  The valence
partons generate classical fields at very small Bjorken $x$.  
The subsequent evolution of these classical fields  
is determined by the  equations of motion.
Therefore, the solution to the equations of
motion is a functional 
of the  valence color charge density (per unit area per unit rapidity) 
in nuclei A and B -- $\rho_A(x_{t})$  and
$\rho_B(x_{t})$, respectively. 
The expectation of any quantity $O$ is 
first expressed in terms of the classical solution
and then averaged over all possible valence charge distributions with 
a weight function. In the McLerran-Venugopalan model, 
a Gaussian ansatz is taken for the weight function. 
Specifically, for a functional 
O($\rho_{A}$,$\rho_{B}$) we have the following average:
\begin{equation}
    \lb O\rb _{\rho}  \equiv 
         \int \left[D\rho_{A}\right] \left[D\rho_{B}\right] 
         O(\rho_{A},\rho_B) 
         \exp\left(  
            -\int d^2 x_t \frac{\mbox{Tr} \left(\rho^2_A(x_t)\right)}{
 \mu^2_{A}(x_t)}  
            -\int d^2 y_t \frac{\mbox{Tr} \left(\rho^2_B(y_t)\right)}
{ \mu^2_{B}(y_t)} 
             \right)  \,.
\end{equation}

At large transverse momentum, the density
of charges is small and the field strength is weak. Therefore
the classical Yang-Mills equations can be linearized and the
solution can be expressed as a functional of
the sources in closed form. The final distribution of gluons can
then be related to the  averages of the gluon field.
The analysis of \cite{KMW}  shows that the distribution
of radiated gluons is given by
\begin{eqnarray}
\label{basic}
   \frac{dN}{dy d^2 k_t} \propto & & \int\,d^2 x_t\, d^2 y_t\, 
e^{i k_t\cdot x_{t}} e^{-i k_t\cdot y_{t} }
\mbox{Tr} \left\langle 
      \left[\partial^{i} \phi_{A}, \partial^{k}\phi_{B} \right](x_{t})
      \left[\partial^{j} \phi_{A}, \partial^{l}\phi_{B} 
\right]^{\dagger}(y_{t})
   \right\rangle _{\rho} \nonumber \\
& &\times (\delta^{ij}\delta^{kl} + \epsilon^{ij}\epsilon^{kl}) \, ,
\end{eqnarray}
where $\phi_{A}$ ($\phi_{B}$) 
is the 2-d Coulomb potential associated  with
the color charge distribution $\rho_{A}$ ($\rho_B$),  
\begin{equation} 
            -\nabla^2_{T}\phi_{A}(x_{T}) = g \rho_{A} \, .
\end{equation}
In what follows, we evaluate the average over $\rho$ 
and and perform the necessary Fourier transforms in order
to determine the gluon distribution to order $\sim (Q_{s} R)^{-2}$.
With this gluon distribution, we determine the azimuthal 
anisotropy, $v_{2}(p_t)$.

\section{Computation of $v_2(k)$ for $k>Q_s$}

The averages over the color sources in Eq. \ref{basic} 
can be performed  most simply in Fourier space. 
Below, we adopt the convention that
repeated spatial indices are integrated over and repeated 
Fourier indices are integrated over with the appropriate
two dimensional measure, $\int \frac{d^2k}{(2\pi)^2}$. In Fourier space, the
average of $O(\rho_A,\rho_B)$ is then
\begin{equation}
    \lb O \rb _{\rho} = 
    \int \left[D\tilde{\rho_{A}}\right] \left[D\tilde{\rho_{B}}\right] 
    O(\rho_{A},\rho_B) 
    \exp\left( -\mbox{Tr} 
           (\tl{\rho}_{A})^*_{k} (C_A)_{kk'} (\tl{\rho}_A)_{k'}   
            -\mbox{Tr} (\tl{\rho}_{B})^*_{k} (C_B)_{kk'} (\tl{\rho}_B)_{k'} \right) 
\end{equation}
where ${(\tl{\rho}_A)_{k}} = e^{ik\cdot x}\rho_A (x)$ is the 
Fourier transform of $\rho_A(x)$ and $(C_A)_{kk'}$ is  given by
\begin{equation} 
	(C_A)_{kk'} = e^{ik\cdot x} \frac{\delta_{xx'}}{\mu^2_{A}(x) } e^{-ik'\cdot x'} \, .
\end{equation}
Then pairwise correlations for nucleus A are given by 
\begin{equation}
  \lb (\tl{\rho_{A}})^{*}_k 
      (\tl{\rho_{A}})_l 
      \rb = (C^{-1}_{A})_{kl} = 
   e^{i (k-l)\cdot x}\mu^{2}_{A}(x) \equiv 
   \tilde{\mu^{2}}_{A}(k-l) \, .
\end{equation}
Fourier transforming each of the $\phi$'s, the gluon distribution of 
Eq. \ref{basic} becomes  
\begin{equation}
\frac{dN}{dy d^2k} \propto
    e^{ik\cdot x_{t}}
    e^{il_{1}\cdot x_{t}}
    e^{il_{2}\cdot x_{t}}
    e^{-ik\cdot  y_{t}}
    e^{-il_{3}\cdot y_{t}}
    e^{-il_{4}\cdot y_{t}}
    \lb 
      \frac{l_{1}}{l_{1}^2} (\tilde{\rho}_A)_{l_1}
      \frac{l_{2}}{l_{2}^2} (\tilde{\rho}_B)_{l_2}
      \frac{l_{3}}{l_{3}^2} (\tilde{\rho}_A)_{l_3}^{*}     
      \frac{l_{4}}{l_{4}^2} (\tilde{\rho}_B)_{l_4}^{*}
    \rb_{\rho} \, .
\end{equation}
Contracting the $\rho_{A}$ variables and the $\rho_{B}$ 
variables we find 
\begin{equation}
\frac{dN}{dy d^2k} \propto
\frac{\delta^{ij}\delta^{kl} + \epsilon^{ij}\epsilon^{kl}}{l_{1}^2 l_{3}^2 (l_{1} + k)^2 (l_{3}+k)^2}
     \, l_{1}^{i}l_{3}^{k} ( l_{1} + k)^{j} (l_{3} + k)^{l}
     (C^{-1}_{A})_{l_1l_3}^* 
     (C^{-1}_{B})_{
                   (l_1+k)
                   (l_3+k)
                  } \, .
\end{equation}
Specializing now to the case where the target and projectile are identical,
 restoring the implicit integration, and 
using $(C_A^{-1})_{kl} = (C_B^{-1})_{kl} = \tilde{\mu^{2}}(k-l)$ we obtain 
\begin{equation}
\label{fundamental}
\frac{dN}{dy d^2k} \propto 
 \int \int 
 \frac{d^2l_{1}}{(2\pi)^2} \frac{d^2l_{3}}{(2\pi)^2}
\frac{\delta^{ij}\delta^{kl} + \epsilon^{ij}\epsilon^{kl}}{l_{1}^2 l_{3}^2 (l_{1} + k)^2 (l_{3}+k)^2}\,
     l_{1}^{i}l_{3}^{k} ( l_{1} + k)^{j} (l_{3} + k)^{l}
     \left|\tilde{\mu^{2}}(l_{1}-l_{3})\right|^{2} \, .
\end{equation}

For the purposes of illustration, we consider a Gaussian model charge 
distribution~\cite{comment}
\begin{eqnarray}
\label{gaussian}
       \mu^2(\vec{x}) &=&   \frac{Q^2}{2\pi R_{x}R_{y}}
                    e^{-\frac{x^2}{2\,R_{x}^2} -\frac{y^2}{2\,R_{y}^2}} \\
       \tilde{\mu^{2}}(\vec{k}) &=& Q^2 e^{-\frac{1}{2} k_{x}^2\,R_{x}^2
                                     -\frac{1}{2} k_{y}^2\,R_{y}^2
                                    }
       \,,
\end{eqnarray}   
where $R_x$ and $R_y$ denote the transverse spatial extent (in the $x$ and 
$y$ direction) of the overlap region of the two nuclei and $Q^2$ is the 
net color charge squared in the overlap region. 

For this model distribution, we see that 
$\left|\tilde{\mu^{2}}(l_1 - l_3) \right|^2$
is a sharply falling function.  Thus 
$\left|\tilde{\mu^{2}}(l_1 - l_3) \right|^2$
is approximately a delta function. Indeed, when 
$\left|\tilde{\mu^{2}}(l_1 - l_3) \right|^2$
is a delta function  Eq. \ref{fundamental} reduces to
the result of \cite{KMW} for the gluon distribution.
We therefore 
approximate the integral over  $\mu^4$ as
\begin{eqnarray}
\label{expansion}
\int \frac{d^2z}{(2\pi)^2} 
\left|\tilde{\mu^{2}}(z) \right|^2 f(z)
 & \approx &
(\tilde{\mu^{4}})^{(0)} f(0) 
 +  
 \frac{1}{2} 
(\tilde{\mu^{4}})^{(2)}_{ij}  
\left(  
   \frac{\partial^2f}{\partial z_i \partial z_j}  
   - \frac{\delta^{ij}}{2}  
     \frac{\partial^2 f}{ \partial_{z_l} \partial_{z_l} }
\right)_{z=0}
\nonumber \\
 &  & +
\frac{1}{2} 
(\tilde{\mu^{4}})^{(2)}_{ij}  
\left(
    \frac{ \delta^{ij}  } { 2 } 
    \frac{ \partial^2 f } { \partial_{z_l} \partial_{z_l} }
\right)_{z=0}  \, .
\end{eqnarray}
In writing the above equation we have assumed that the 
charged distribution 
$\left| \tilde{\mu^{2}}(z) \right|^2$
is symmetric in $z$  and  defined 
the moments of this distribution 
as
\begin{eqnarray}
(\tilde{\mu^{4}} )^{(0)}  
 & \equiv &
\int \frac{ d^2z }{ (2\pi)^2 }
\left| 
      \tilde{\mu^{2}}(z) \right|^2
\\
(\tilde{\mu^{4}} )^{(2)}_{ij}  
 & \equiv &
\int \frac{d^2z}{(2\pi)^2}
z_{i} z_{j} 
\left|\tilde{\mu^{2}}(z) \right|^2 \, .
\end{eqnarray}
For the problem at hand, we have 
\begin{equation}
\label{function}
\frac{dN}{dy d^2k} \propto 
\int \frac{d^2z}{(2\pi)^2} 
\left|\tilde{\mu^{2}}(z) \right|^2
f(k,z)
\end{equation}
where $f(k,z)$ is given by
\begin{equation}
f(k,z) \equiv
\int \frac{d^2p}{(2\pi)^2}
\frac{\delta^{ij}\delta^{kl} + \epsilon^{ij}\epsilon^{kl}}
{p^2 (p-z)^2 (p + k)^2 (p - z + k)^2}
    p^{i}(p-z)^{k} (p + k)^{j} (p - z + k)^{l} \,.
\end{equation}
We may write down the
general form of  the second derivatives of $f(k,z)$
\begin{eqnarray}
\label{derivs}
\left(  
     \frac{\partial^2f(k,z)}{\partial z_i \partial z_j}  
   - \frac{\delta^{ij}}{2}  
     \frac{\partial^2 f(k,z)}{ \partial_{z_l} \partial_{z_l} }
\right)_{z=0}  &\equiv&   
f^{(2)}(k^2)  (\frac{k_ik_j}{k^2} - \frac{\delta^{ij}}{2})  \\
\left( 
     \frac{\partial^2 f(k,z)}{ \partial_{z_l} \partial_{z_l} }
\right)_{z=0} & \equiv &
g^{(2)}(k^2) \, .
\end{eqnarray} 
Now the elliptic flow  parameter $v_{2}$ is given by
\begin{equation}
\label{v2define}
v_{2}(y,k) \equiv 
\frac {
   \int \frac{d\phi}{2\pi}  \cos(2\phi)
   \frac{dN}{dy\, k\,dk\, d\phi} 
}{
   \int \frac{d\phi}{2\pi} 
   \frac{dN}{dy\, k\,dk\, d\phi}  
} \,.
\end{equation}
Substituting Eq. \ref{expansion} and Eq. \ref{derivs}  into 
this definition of $v_{2}$, we obtain to leading order in $(Q_{s}R)^{-1}$
\begin{equation}
v_{2}(y,k) =  
\frac{
   (\mu^4)^{(2)}_{xx} - 
   (\mu^4)^{(2)}_{yy}   
}{
   8 (\mu^4)^{(0)}
}
\frac{ 
   f^{(2)} (k^2) 
}{
   f(k,z=0)
} \,.
\end{equation}
$f(k,z=0)$ and $f^{(2)}(k^{2})$ are calculated  in 
appendix A. The results are logarithmically
divergent. As discussed in \cite{KMW}, this 
is a consequence of the weak field expansion; higher order
non-linear corrections cut off the behavior in the
infra-red at a scale $\alpha_{s} \mu$. To leading log accuracy
we find
\begin{eqnarray} 
f(k,z=0) &\approx& \frac{1}{(2\pi)} 
           \frac{1}{k^2}
           \log\left(\frac{k^2}{(\alpha_{s} \mu)^2}\right)  \\
f^{(2)}(k^2) &\approx&  \frac{1}{ (2\pi)} 
           \frac{4}{k^4}
           \log\left(\frac{k^2}{(\alpha_{s} \mu)^2}\right) \, .
\end{eqnarray}
With these expressions we obtain our final result for $v_{2}$ 
\begin{equation}
\label{v2final}
v_{2}(y,k) \approx  
\frac{1}{2 k^2}
\frac{
   (\mu^4)^{(2)}_{xx} - 
   (\mu^4)^{(2)}_{yy}   
}{
   (\mu^4)^{(0)}
} \,.
\end{equation}
It is remarkable (see the appendix) that there are no power or even 
logarithmic divergences from the ratios of 
integrals in our expression for $v_2$. 
For the Gaussian distribution of Eq. \ref{gaussian} the moments of the charge 
distribution are easily calculated. 
The resulting elliptic flow is 
\begin{equation}
v_{2}(y,k) \approx  
\frac{1}{4 k^2}(\frac{1}{R^2_{x}} - \frac{1}{R^2_{y}} )\, .
\label{result}
\end{equation}
The contribution of momenta $k>Q_s$ to the integrated $v_2$ (Eq.~\ref{eq1})
is 
\begin{equation}
v_2 = {\int_{Q_s}^\infty k\,dk\, v_2(k) \frac{dN}{d^2 k} \over
{\int_{Q_s}^\infty d^2 k\,\frac{dN}{d^2 k}}} \propto \frac{1}{Q_s^2}\left( 
\frac{1}{R_x^2}-\frac{1}{R_y^2}\right) \, .
\end{equation}
For realistic values of $Q_s\sim 1$ GeV for RHIC and $R_x\sim 3$ fm for a
peripheral collision, this contribution is a fraction of a percent. The 
significant contribution of classical fields to $v_2$ comes from momenta 
$k< Q_s$. Interestingly, our results suggest that $v_2$ for $k>Q_s$ is 
generated very early. The spatial and temporal components decouple 
already at proper time $\tau=0$. The temporal contributions are 
Bessel functions $J_0(k\tau)$ and $J_1(k\tau)$ which can be linearized 
into plane waves for times $\tau<1/Q_s$ when $k>>Q_s$. 

\section{Summary}

In this paper, we have computed analytically the contribution to the 
azimuthal anisotropy ($v_2(k)$) from classical fields at large transverse 
momenta, $k^2 > Q_s^2$.  We find that the azimuthal anisotropy has the 
simple form shown in Eq.~\ref{result} for a Gaussian charge distribution. 
Clearly, this behavior disagrees with the data and another mechanism must 
be found to explain the large momentum behavior. 
Non-flow azimuthal correlations may provide a natural 
explanation \cite{Yuri}. An interesting 
feature of our result is that the 
temporal dependence of the perturbative classical fields decouples already 
at proper time $\tau=0$~\cite{KMW}. 
Thus as Eq.~\ref{basic} suggests, the 
momentum anisotropy is generated at very early times in the collision. 
This analysis may thus explain why (in numerical simulations of the Yang-Mills 
equations~\cite{Yasushi}) a significant anisotropy is seen at early times 
after the collision.

\section{Acknowledgements}
We would like to thank D. Kharzeev, Y. Kovchegov, A. Krasnitz, L. McLerran, 
Y. Nara and K. Tuchin for useful discussions. R.V and D.T.'s work was 
supported by DOE Contract No. DE-AC02-98CH10886. 
R.V would also like to acknowledge RIKEN-BNL for support.

\appendix
\section{Evaluation of Integrals}
\label{appI}

To complete the evaluation of $v_{2}$ we expand 
Eq. \ref{function} as a function of $z$ and extract the coefficients
of the symmetric traceless tensor  structures. We
find after some algebra
\begin{eqnarray}
    f(k,z=0) &=&  
    \int \frac{d^{2}p}{(2\pi)^2} 
           \frac{1}{p^2 (k+p)^2}   \\
\label{integrals}
    f^{(2)}(k^2) (\frac{k_{i}k_{j}}{k^2} - \frac{\delta_{ij}}{2}) &=& 
                   (I^{1})_{ij}
                  +(I^{2})_{ij} 
                  +(I^{3})_{ij} 
\end{eqnarray}
where the  integrals  are given by
\begin{eqnarray}
   (I^{1})_{ij} &=& \int \frac{d^{2}p}{(2\pi)^2} 
           \frac{2}{p^2 (k+p)^6}  
           \left(k_i k_j - k^2      \frac{\delta_{ij}}{2}\right) \\
   (I^{2})_{ij} &=& \int \frac{d^{2}p}{(2\pi)^2} 
           \frac{4}{p^2 (k+p)^6} 
           \left({k_i p_j + p_i k_j\over 2} - 
p \cdot k \frac{\delta_{ij}}{2}\right) \\
   (I^{3})_{ij} &=& \int \frac{d^{2}p}{(2\pi)^2}  
           \left(\frac{2}{p^2 (k+p)^6} +  \frac{2}{p^6(p+k)^2} \right)
           \left(p_i p_j -  p^2     \frac{\delta_{ij}}{2}\right)  \, .
\label{sums}
\end{eqnarray}
$f(k,z=0)$ is logarithmically divergent at $p=0$ and $\vec{p}=-\vec{k}$. It 
is identical to the expression obtained previously in 
\cite{KMW,GyulassyMcLerran}. 
As discussed there and shown explicitly in Ref.~\cite{KV} 
this infra-red divergence is an artifact of the weak field expansion. To
leading log accuracy we find
\begin{equation}
    f(k,z=0) \approx 
    \frac{1}{(2\pi)}
    \frac{1}{k^2} \log\left(\frac{k^2}{(\alpha_{s}\mu)^2}\right)  \, .
\end{equation}
Consider now the sum of the integrals in Eqs.~\ref{sums}. While each of 
these individually have power divergences, taken together, the result is 
again only logarithmically divergent. Note, as previously, the integrals 
are divergent at $p=0$ and $\vec{p}=-\vec{k}$ and one obtains logarithmic 
contributions from both regions to the final result. For illustration, 
consider the contribution to the sum from the divergence around $p=0$. 
$(I^{1})_{ij}$ is logarithmically divergent, $(I^{2})_{ij}$ is finite, and  
$(I^{3})_{ij}$ appears power divergent. However, $(I^{3})_{ij}$ 
is actually only logarithmically divergent upon integration over the
angle  between $p$ and $k$.
Contracting both sides of Eq. \ref{integrals} with 
$2({k_{i}k_{j}\over k^2} - \frac{\delta_{ij}}{2})$ and keeping only 
logarithmically divergent terms we obtain the contribution of the 
$p=0$ divergence to $(I^{1})_{ij}$ and $(I^{3})_{ij}$ respectively to be 
\begin{eqnarray}
2(I_{1})_{ij} 
\left(\frac{ k^{i} k^{j} }{k^2} 
   - \frac{\delta^{ij}}{2} \right)_{\rm{|p|\approx \alpha_s\mu}} 
  &=& k^{2} \int \frac{d^2p}{(2\pi)^2} 
    \frac{2}{p^2 (k+p)^6}  
    \approx 
    \frac{1}{(2\pi)}
    \frac{1}{k^4} \log\left(\frac{k^2}{(\alpha_{s}\mu)^2}\right)\nonumber \\
2 (I_{3})_{ij} 
\left(\frac{ k^{i} k^{j} }{k^2} 
   - \frac{\delta^{ij}}{2} \right)_{\rm{|p|\approx \alpha_s\mu}} 
  &=&  
  \int \frac{d^2p}{(2\pi)^2} 
    \frac{2\,p^2 \cos(2\phi_{pk})}{p^6 (k+p)^2} 
    \approx 
    \frac{1}{(2\pi)}
    \frac{1}{k^4} \log\left(\frac{k^2}{(\alpha_{s}\mu)^2}\right) 
\label{diverge}
\end{eqnarray}
where $\phi_{pk}$ is the angle between $p$ and $k$. A detailed analysis 
shows that the 
contribution to {\it the sum} $(I^{1})_{ij}+(I^{2})_{ij}+(I^{3})_{ij}$ 
from the logarithmic divergence around $\vec{p}=-\vec{k}$ is identical 
to the sum of the divergent pieces in Eqs.~\ref{diverge}. 

Substituting the results of the above analysis 
into the expression for $f^{(2)}(k^2)$ we find 
\begin{equation}
    f^{(2)}(k^{2})
    = 
    \frac{1}{(2\pi)}
    \frac{4}{k^4} \log\left(\frac{k^2}{(\alpha_{s}\mu)^2}\right)  \, .
\end{equation}
With these expressions for $f(k,z=0)$ and $f^{(2)}(k^2)$, 
the final result for $v_{2}$ given in Eq. \ref{v2final} follows.


\begin{thebibliography}{9}

\bibitem{Expt1}
   STAR Collaboration, K.H. Ackermann {\it et al.}, 
   Phys.\ Rev.\ Lett.\ {\bf 86}, 402 (2001);
   STAR Collaboration, C.~Adler {\it et al.}, Phys.\ Rev.\ Lett.\
   {\bf 87}, 182301 (2001).
\bibitem{CERN}
   NA49 Collaboration, H.~Appelsh\"{a}user {\em et al.},
   Phys.\ Rev.\ Lett.\  {\bf 80}, 4136  (1998);
   A.M. Poskanzer and S.A.~Voloshin for the NA49 Collaboration,
   Nucl.\ Phys.\ A{\bf 661}, 341c  (1999).
\bibitem{Hydro1}
   D.~Teaney, J.~Lauret, and E.V.~Shuryak, 
   Phys.\ Rev.\ Lett. {\bf 86}, 4783 (2001); 
   D.~Teaney, J.~Lauret, and E.V.~Shuryak,  nucl-th/0110037.
\bibitem{Hydro2}P.F. Kolb, P.Huovinen, U. Heinz, H. Heiselberg,
   Phys.\ Lett.\ B {\bf 500}, 232 (2001);
   P. Huovinen, P.F. Kolb,  U. Heinz, H. Heiselberg,
   Phys.\ Lett.\ B {\bf 503}, 58 (2001).
\bibitem{Expt2}
   R.J.M.~Snellings for the STAR Collaboration, 
   Nucl.\ Phys.\ A{\bf 698}, 611c  (2002).
\bibitem{Blast}
   P.J. Siemens and J.O. Rasmussen, Phys.\ Rev.\ Lett.\ {\bf 42}, 880 (1979);
   E. Schnederman, J. Sollfrank, U.Heinz, Phys.\ Rev.\  C{\bf 48}, 2462 (1993);
   P. Huovinen, P.F. Kolb, U. Heinz, Nucl.\ Phys.\ A{\bf 698}, 475 (2002). 
\bibitem{GVW}
   M.~Gyulassy, I.~Vitev, X.N.~Wang, Phys.\ Rev.\ Lett.\ {\bf 86}, 2537 (2001).
\bibitem{Molnar}
   D. Molnar and M. Gyulassy, Nucl.\ Phys.\ A{\bf 697}, 495 (2002);
   D. Molnar and M. Gyulassy, Nucl.\ Phys.\ A{\bf 698}, 379 (2002). 
\bibitem{MV}L.~McLerran and R.~Venugopalan,
Phys.\ Rev.\ D {\bf 49}, 2233 (1994); {\bf 49}, 3352 (1994); D {\bf 50}, 2225 (1994); D {\bf 59}, 094002 (1999); J.~Jalilian-Marian, A.~Kovner, 
L.~McLerran and H.~Weigert,
Phys.\ Rev.\ D {\bf 55}, 5414 (1997); Yu.~V.~Kovchegov,
Phys.\ Rev.\ D {\bf 54}, 5463 (1996).
\bibitem{KMW}A.~Kovner, L.~McLerran and H.~Weigert,
Phys.\ Rev.\ D {\bf 52}, 3809 (1995); D {\bf 52}, 6231 (1995).
\bibitem{KR}Yu.~V.~Kovchegov and D.~H.~Rischke,
Phys.\ Rev.\ C {\bf 56}, 1084 (1997).
\bibitem{KV}A.~Krasnitz and R.~Venugopalan,
Nucl.\ Phys.\ B {\bf 557}, 237 (1999); 
Phys.\ Rev.\ Lett.\ {\bf 84}, 4309 (2000); {\it ibid.} {\bf 86}, 1717 (2001);
A.~Krasnitz, Y.~Nara and R.~Venugopalan, 
Phys.\ Rev.\ Lett.\  {\bf 87}, 192302 (2001).
\bibitem{GLR}L.V.~Gribov, E.M.~Levin, M.G. Ryskin, 
Phys.\ Rept.\ {\bf 100}, 1 (1983).
\bibitem{GyulassyMcLerran} M.~Gyulassy and L.~McLerran,
Phys.\ Rev.\ C {\bf 56}, 2219 (1997).
\bibitem{Mueller}A. H. Mueller, Nucl.\ Phys.\ B {\bf 572}, 227 (2000).
\bibitem{KN}D.~Kharzeev and M.~Nardi, 
Phys.\ Lett.\ B {\bf 507}, 121 (2001).
\bibitem{KL} D.~Kharzeev and E.~Levin
nucl-th/0108006 .
\bibitem{Juergen}L.~McLerran and J.~Schaffner-Bielich,
hep-ph/0101133; J.~Schaffner-Bielich, D.~Kharzeev,
L.~D.~McLerran and R.~Venugopalan, nucl-th/0108048 .
\bibitem{comment}In general, one will obtain a non-zero contribution 
to $v_2(p_t)$ in the classical approach as long as a) the nuclei are 
finite and b) anisotropic.
\bibitem{Yasushi}A.~Krasnitz, Y.~Nara and R.~Venugopalan, to be published.
\bibitem{Yuri}Y. V. Kovchegov and K. Tuchin, hep-ph/0203213.

\end{thebibliography}
\end{document}